\documentclass[aps,prl,twocolumn,showpacs,superscriptaddress]{revtex4}
\usepackage{graphicx}
\usepackage{float}
\usepackage{amsmath}
\usepackage{amssymb}
\usepackage{color}
\usepackage{braket}
\usepackage{hyperref}
\graphicspath{{converted_graphics/}}

\begin{document}
\title{Real-time Phase Conjugation of Vector Vortex Beams}

\author{A. G. de Oliveira}
\affiliation{Departamento de F\'isica, Universidade Federal de Santa Catarina, Florian\'opolis, SC, 88040-900, Brazil}

\author{M. F. Z. Arruda}
\affiliation{Departamento de F\'isica, Universidade Federal de Santa Catarina, Florian\'opolis, SC, 88040-900, Brazil}
\affiliation{Instituto Federal do Mato Grosso, Sorriso, MT, 78890-000, Brazil}

\author{W. C. Soares}
\affiliation{Universidade Federal de Alagoas, Campus Arapiraca, Arapiraca, AL, 57309-005, Brazil}

\author{S. P. Walborn} 
\affiliation{Instituto de F\'isica, Universidade Federal do Rio de Janeiro, Rio de Janeiro, RJ, 21945-970, Brazil}

\author{R. M. Gomes}
\affiliation{Instituto de F\'isica, Universidade Federal de Goi\'as, Goi\^ania, GO, 74690-900, Brazil}

\author{R. Medeiros de Ara\'ujo}
\affiliation{Departamento de F\'isica, Universidade Federal de Santa Catarina, Florian\'opolis, SC, 88040-900, Brazil}

\author{P. H. Souto Ribeiro} 
\affiliation{Departamento de F\'isica, Universidade Federal de Santa Catarina, Florian\'opolis, SC, 88040-900, Brazil}

\date{\today}
\begin{abstract}
Vector vortex beams have played a fundamental role in the better understanding of coherence and polarization. They are described by spatially inhomogeneous polarization states, which present a rich optical mode structure that has attracted much attention for applications in optical communications, imaging, spectroscopy and metrology. However, this complex mode structure can be quite detrimental when propagation effects such as turbulence and birefringence perturb the beam.  Optical phase conjugation has been proposed as a method to recover an optical beam from perturbations.   Here we demonstrate full phase conjugation of vector vortex beams using three-wave mixing.  Our scheme exploits a fast non-linear process that can be conveniently controlled via the pump beam.   Our results pave the way for sophisticated, practical applications of vector beams. 
\end{abstract}
\maketitle
%
{Vector vortex beams  are at the heart of many applications in optics, due to their more complex optical mode structure, described by anisotropic polarization states that are not spatially homogeneous in the transverse plane. They can provide crucial benefits in important applications, such as increased transmission rates in classical and quantum optical communications in free space \cite{dambrosio12,Zhao15,Farias15,Milione15,Milione15A,zhang16,Li16,Ndagano18}, as well as in optical fibers \cite{Gregg15,Ndagano15}.  Unique approaches for enhanced-resolution imaging \cite{Biss06,Yoshida19} and remote optical sensing \cite{dambrosio13b,Toppel2014,Berg-Johansen15} have been demonstrated with vector beams.  The radial vector beam has tighter focusing properties that have been used in nanoscale optical imaging \cite{Zhan09,Chen13} and laser cutting \cite{Niziev1999}, while it's radial polarization boosts efficiency in tip-enhanced near-field spectroscopy and imaging \cite{Schultz09,Kazemi-Zanjani13,Lu18}.  While vector beams show much promise for interesting applications, the spatial mode structure of these beams can make them quite prone to distortions caused by turbulence or diffusive processes, while the polarization properties are degraded by dynamical optical birefringence, which can appear during propagation in optical fibers, for example.
\par
As is well known, optical wavefronts suffer distortion during propagation in anisotropic media or multiple light scattering \cite{Fisher2012,MacDonald1988,Boyd1989}, preventing focusing and imaging.
One method to correct these perturbations is through phase conjugation, as was first reported in 1970 by Zel’dovich et al. \cite{Zel1995}.  They observed that light emerging in reverse from a distorted glass displayed a spatial profile that was free from the distortions of the diffusive medium, after back scattering on a carbon disulfide (CS$_2$) gas cell, which functioned as a phase conjugating mirror (PCM). The PCM essentially realized the time reversal of the optical field, which allowed the distortions to be corrected when the field back-propagated through the medium.  Wavefront correction by optical phase conjugation has been used to improve high-resolution imaging \cite{Levenson1980} and to construct laser oscillators \cite{McFarlane1983,Gower1982}. More recently, it was used in biological tissue to control waves in space and time \cite{Mosk2012,Yaqoob2008}, and is viewed as an essential component for  long distance optical fiber communications with ultra-high bit-rate \cite{Set98}. }
\par
Various techniques have been explored to produce a phase-conjugate beam, including degenerate four-wave mixing \cite{Yariv1977,Boyd1989,Damzen1996,Gunter2007}, backward stimulated Brillouin, Raman, and Rayleigh-wing or Kerr scattering \cite{Zel1995}, as well as single- or multiphoton-pumped backward stimulated emission (lasing) \cite{He2002}.  Though most experiments have focused on spatial properties of the field, some have concerned phase conjugation of isotropic polarization states\cite{Boyd1989}.   Phase conjugation of anisotropic vector fields was realized in Refs. \cite{Qian14_OL1,Qian14_OL2} in photorefractive media, however the required exposure time of 250s is typically much too long a time scale to correct wavefront distortions accrued during propagation.
\par
Here we demonstrate phase conjugation of vector beams using Stimulated Parametric Down-Conversion (StimPDC), a three-wave mixing process \cite{Wang90,Wang91}.  StimPDC has previously been used to realize phase conjugation in the transverse spatial profile \cite{PHSR2001,Arruda2018}, including light beams possessing orbital angular momentum \cite{Caetano2002,Oliveira2019}.   We employ an arrangement of non-linear crystals that produces an idler beam that is the phase conjugate of the seed beam, in both the polarization and spatial degrees of freedom.  StimPDC presents a number of advantages as a phase conjugation process.  First, the process is fast, occurring essentially in real time, as the emission is stimulated and therefore follows the seed frequency. This renders StimPDC as a promising technique for correcting dynamic distortions of optical beams in general. Second, we show that the polarization state of the pump beam provides convenient control parameters, allowing one to rotate the polarization states affected by phase conjugation and switch conjugation on and off.  
\paragraph{\bf{Phase Conjugation}}\par
\begin{figure}
    \centering
    \includegraphics[width=\columnwidth]{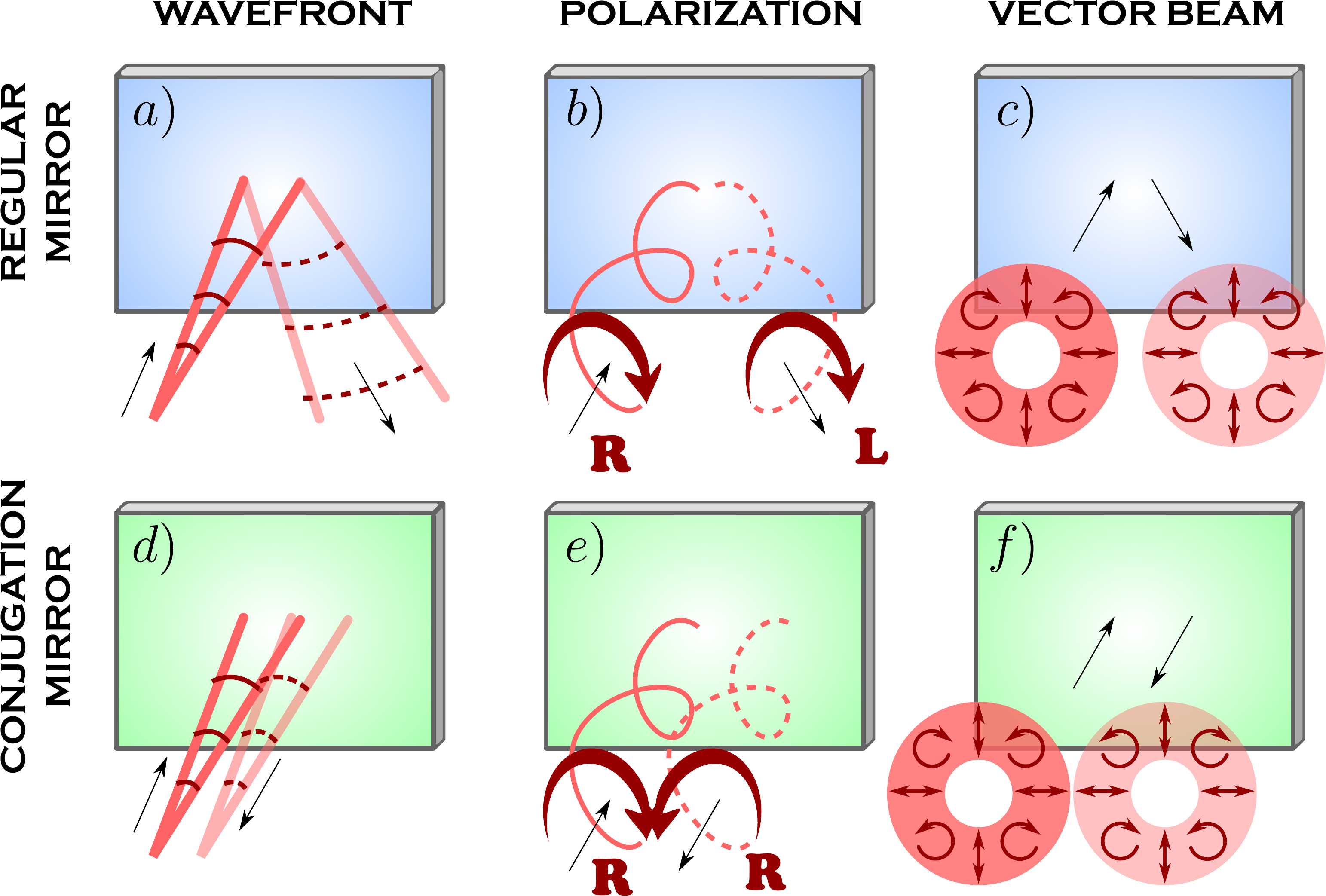} 
     \caption{Comparison between an ordinary mirror and a phase conjugation mirror in terms of wavefront, polarization and vector beam. Black arrows indicate direction and sense of propagation.}
       \label{fig:PCM}
\end{figure}

 Let us recall the differences between an ordinary mirror and a ideal PCM. While the ordinary mirror reverses only the normal component of the wavevector, the ideal PCM not only reverses the entire wavevector $\vec{k}$, but also conjugates the field amplitude (phase profile: $\phi(\vec{r}_\perp)\rightarrow -\phi(\vec{r}_\perp)$; polarization: $\vec{\varepsilon}\rightarrow\vec{\varepsilon}^{\,*}$). Fig. \ref{fig:PCM} a) shows a diverging beam incident on an ordinary mirror.  After reflection, it continues diverging.  On the other hand, the same beam incident on a PCM converges backwards towards the source, as illustrated in Fig.\ref{fig:PCM} d). This is equivalent to time reversal of the field. Thus, reflection upon a PCM is not specular, meaning that the angle of incidence is not equal to the angle of reflection. In terms of the polarization of the light beam,  reflection of a circularly polarized field from an ordinary mirror and a PCM is illustrated in Fig. \ref{fig:PCM} b) and Fig. \ref{fig:PCM} e), respectively.  The reflected beam changes its circular polarization state from $R$($L$) to $L$($R$) when reflected from the ordinary mirror, but remains unchanged when reflected from the PCM, since the polarization is conjugated and also the wave vector is reversed. We can use this picture to infer what happens to a vector beam, as shown in Figs. \ref{fig:PCM} c) and f). There is no difference between specular and PCM reflection for linear polarization components. However, for circular components they are opposed. The PCM reflects an arbitrary and unknown input state conjugating all polarization states present in the transverse spatial profile. This cannot be achieved using specular devices without the previous knowledge of the input state. In this sense, phase conjugation can be used to restore arbitrary vector beams perturbed by propagation through some anisotropic and/or birefringent medium.

\par
  StimPDC differs from the ideal PCM considered in Fig. \ref{fig:PCM} in that it is a forward process, in which the pump, seed and idler all propagate in the forward direction.  The generated idler beam displays the conjugate of the transverse phase profile of the seed laser beam (signal). In this process, the wavevector $\vec{k}$ is not reversed but rather reflected in the transverse plane.

\paragraph{\bf{Spontaneous three-wave mixing in two crystals}}
\label{sec:theory}
\begin{figure}
\centering
   \includegraphics[width=0.75\columnwidth]{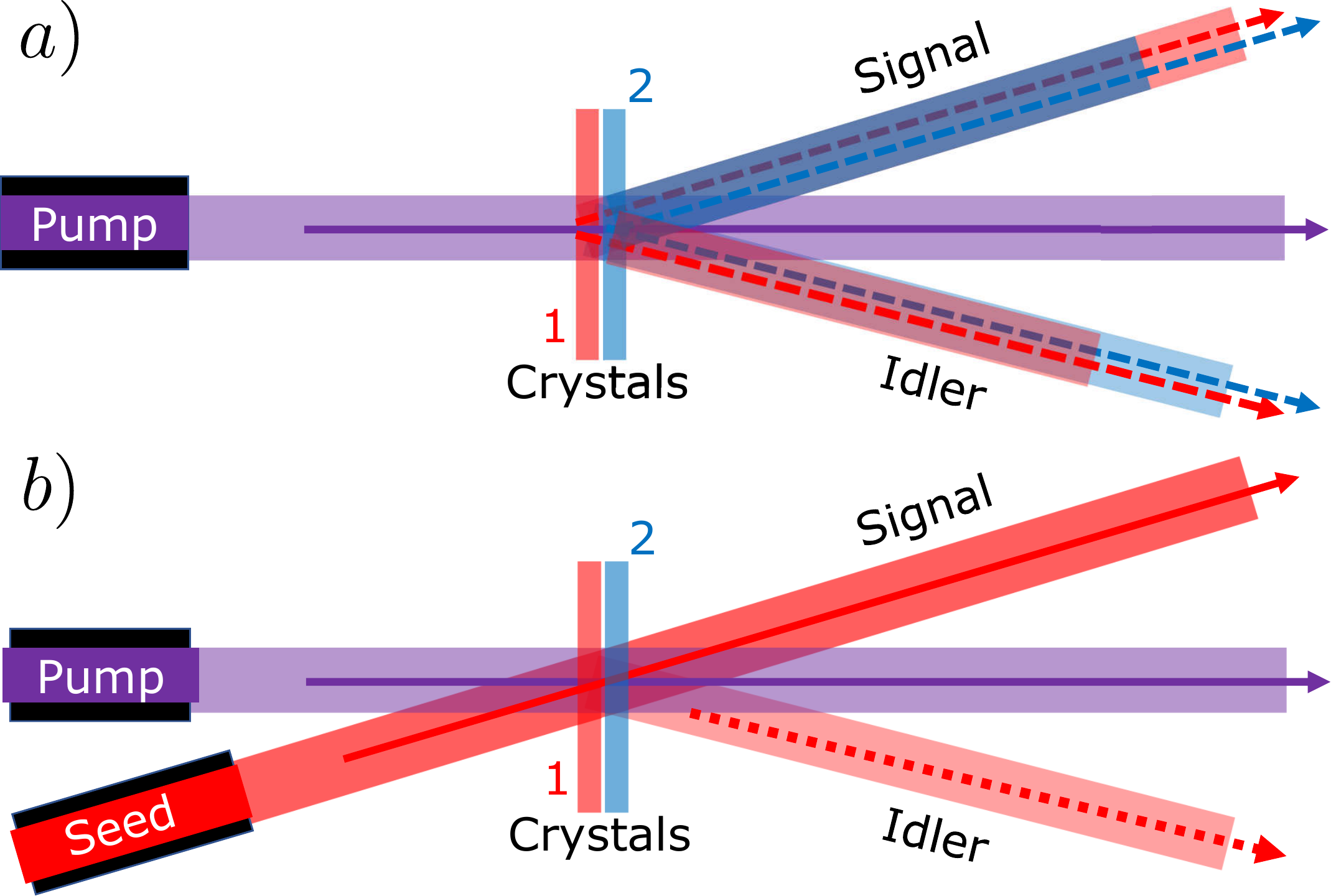}
   \caption{a) Two-crystal type-I parametric down-conversion source for the generation of polarization Bell-states. b) Stimulated parametric down-conversion for two-crystal source configuration.}
   \label{fig:2crystals}
\end{figure}
We first demonstrate a polarization-conjugation device using StimPDC based on a two-crystal geometry, inspired by  a spontaneous parametric down conversion (SPDC) source of polarization-entangled photons \cite{Kwiat1999} as illustrated in Fig. \ref{fig:2crystals} a).  The optical axes of the crystals are crossed, so that one of them emits vertically-polarized pairs of photons and the other emits horizontally-polarized ones. Here we focus on the polarization degree of freedom, so that pump, signal and idler fields will be treated as ideal monochromatic plane waves, which can be achieved using spatial and spectral filters. 

Let us consider that the pump beam is described by the general polarization state
\begin{equation}
\ket{\vartheta_p,\varphi_p}=\cos{\frac{\vartheta_p}{2}}\ket{H}+e^{{i(\varphi_p-\Phi)}}\sin{\frac{\vartheta_p}{2}}\ket{V},
\label{eq:p-state}
\end{equation}
where the extra phase $-\Phi$ is added to offset the phase accrued between the SPDC crystals. If light emitted by the two crystals is spectrally and spatially indistinguishable, the polarization state of the SPDC pair will be completely determined by the pump beam:
\begin{equation}
\ket{\psi}_{s,i}=\cos{\frac{\vartheta_p}{2}}\ket{V}_s\ket{V}_i+e^{i\varphi_p}\sin{\frac{\vartheta_p}{2}}\ket{H}_s\ket{H}_i.
\label{eq:SPDC-state}
\end{equation}
It will be convenient to define the orthonormal states
\begin{align}
\begin{split}
\ket{\pm\vartheta_s,\varphi_s}&=\cos{\frac{\vartheta_s}{2}}\ket{H} \pm e^{i\varphi_s}\sin{\frac{\vartheta_s}{2}}\ket{V},
\label{eq:pol1}
\end{split}
\end{align}
so that the SPDC state (\ref{eq:SPDC-state}) can be rewritten as
\begin{equation}
    \ket{\psi}_{s,i}=   \ket{+\vartheta_s,\varphi_s}\ket{\alpha}+\ket{-\vartheta_s,\varphi_s}\ket{\beta},
    \label{eq:Bellgeneral}
\end{equation}
where
\begin{align}
\begin{split}
 \ket{\alpha}&=\sin{\frac{\vartheta_p}{2}}\cos{\frac{\vartheta_s}{2}}\ket{H}+e^{-i(\varphi_p+\varphi_s)}\cos{\frac{\vartheta_p}{2}}\sin{\frac{\vartheta_s}{2}}\ket{V} ,
\\
\ket{\beta}&=\sin{\frac{\vartheta_p}{2}}\sin{\frac{\vartheta_s}{2}}\ket{H}-e^{-i(\varphi_p+\varphi_s)}\cos{\frac{\vartheta_p}{2}}\cos{\frac{\vartheta_s}{2}}\ket{V}.
\label{eq:alpha-beta}
\end{split}
\end{align}

The state (\ref{eq:Bellgeneral}) represents the possible polarization-state outputs for spontaneous emission from the two-crystal source. 

\paragraph{Phase conjugation in stimulated emission}

In StimPDC, a seed beam stimulates the emission--say--in the signal mode, also stimulating emission in the idler mode, since the down-converted photons are produced in pairs. In order to maximize efficiency,  the optical mode of the seed laser must match the desired signal mode. 
\begin{figure}[h]
\centering
   \includegraphics[width=\columnwidth]{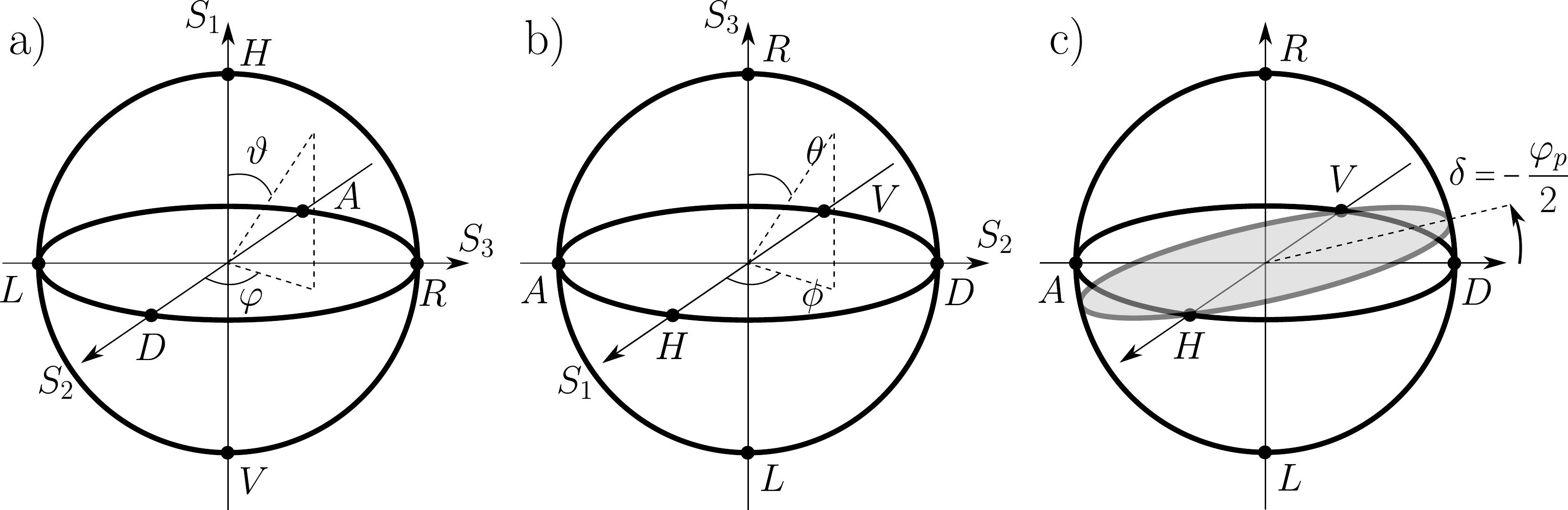}
   \caption{Poincar\'e sphere with N/S poles given by a) H/V polarization, b) R/L polarization and c) Signal and idler polarizations are mirror images through the $\delta$-rotated conjugation plane in StimPDC when $\vartheta_p=\pi/2$.}
   \label{fig:poincare}
\end{figure}

Let us consider the two-crystal configuration shown in Fig. \ref{fig:2crystals} b), 
where the seed beam is prepared with polarization state $\ket{\vartheta_s,\varphi_s}$, which will stimulate signal photons also in polarization $\ket{\vartheta_s,\varphi_s}$. As a consequence of the polarization correlations described in Eq. (\ref{eq:Bellgeneral}), which are imposed by the two-crystal configuration and phase matching conditions, this enhances the emission of idler photons with polarization $\ket{\alpha}$ given in Eq. (\ref{eq:alpha-beta}). If the stimulated emission rate in the idler beam is high enough, the spontaneous emission, which is always present, can be neglected, and the signal and idler fields are approximately described by coherent states. 

It is illustrative to describe the polarization states using the Poincar\'e sphere, which has the circular polarizations at the poles and is a parametrization of the last three Stokes' parameters in spherical coordinates $(\rho,\theta,\phi)$, as illustrated in Fig. \ref{fig:poincare} b). The polar and azimuthal angles $\theta$ and $\phi$ are used to describe polarization in the circular $\{\ket{R}, \ket{L}\}$ basis, similarly to how $\vartheta$ and $\varphi$ describe polarization in the $\{\ket{H}, \ket{V}\}$ basis (see equations (\ref{eq:p-state}) and (\ref{eq:pol1}) and Fig. \ref {fig:poincare} a)). While the latter is a privileged basis for describing the coupling of light with the each crystal, the former has been conventionally used in the Poincar\'e representation and is also convenient for the geometrical interpretation of our results, as introduced below.
Henceforth, we will refer to the Stokes vector $\vec{S}=(S_1, S_2, S_3)^T$, omitting dependence on the intensity parameter $S_0$, which is not relevant to our analysis. On the Poincar\'e spheres, the polarization state $\ket{\vartheta_\mu,\varphi_\mu}$ has Stokes vector
\begin{equation}
    \Vec{S_\mu} =\left( 
    \begin{matrix}
        \cos{\vartheta_\mu} \\
        \sin{\vartheta_\mu}\cos{\varphi_\mu} \\
        \sin{\vartheta_\mu}\sin{\varphi_\mu}
    \end{matrix}
    \right)=\left( 
    \begin{matrix}
        \sin{\theta_\mu}\cos{\phi_\mu} \\
        \sin{\theta_\mu}\sin{\phi_\mu} \\
        \cos{\theta_\mu} \\
    \end{matrix}
    \right),
    \label{eq:p-s-stokes}
\end{equation}
where $\mu=p,\,s,\,i$ may represent pump, signal/seed or idler.

In two-crystal StimPDC, the idler polarization state $\ket{\alpha}$ is completely defined by the Stokes parameters of the pump and seed beams. Indeed, from equation (\ref{eq:alpha-beta}), one can show that 
\begin{align}
    \Vec{S_i} =\frac{1}{2}\left( 
    \begin{matrix}
        S_{s,1}-S_{p,1} \\
        S_{p,2}S_{s,2}-S_{p,3}S_{s,3} \\
        -S_{s,2}S_{p,3}-S_{p,2}S_{s,3}
    \end{matrix}
    \right).
    \label{eq:idler-stokes}
\end{align}

The above equation is our first theoretical result and we will use it to analyze the polarization (vector) conjugation of the signal beam. Polarization conjugation in terms of Stokes vectors is defined as $\Vec{S}^*=(S_1,S_2,-S_3)^T$ \cite{Boyd1989}. Comparing this definition with Eq. (\ref{eq:idler-stokes}), we see that, if the pump beam polarization is linear diagonal ($D$) $\vec{S}_p=(0,1,0)^T$, 
the idler is given by
\begin{equation}
    \Vec{S_i}=(S_{s,1},S_{s,2},-S_{s,3})^T=\Vec{S_s}^*, 
\end{equation}
equal to the conjugate of the polarization of the seed beam. As the conjugation flips only the sign of the third Stokes vector component, we can interpret this result geometrically in the following way: on the Poincar\'e sphere of polarization, signal and idler are mirror images of each other by reflection on the equatorial plane. 

Note that choosing the pump beam with linear anti-diagonal ($A$) polarization $\vec{S}_p=(0,-1,0)^T$ gives
\begin{equation}
    \Vec{S_i}=(S_{s,1},-S_{s,2},S_{s,3})^T,
\end{equation}
so that signal and idler are mirror images of each other by reflection on the vertical plane $S_2=0$. In fact, there are many interesting intermediate polarization transformations that can be implemented by choosing different pump beam polarization states.  The particular case of the pump beam containing equal amounts of $H$ and $V$ polarizations in a coherent superposition ($\vartheta_p=\pi/2$ or, equivalently, $S_{p,1}=0$) is especially interesting, since the coupling to both crystals is the same.  In this case, we may rewrite equation (\ref{eq:alpha-beta}) as
\begin{equation}
    \ket{\alpha}=\cos{\frac{\vartheta_s}{2}}\ket{H}+e^{i\left(2\delta-\varphi_s\right)}\sin{\frac{\vartheta_s}{2}}\ket{V},
\end{equation}
where $\delta=-\varphi_p/2$. The relative-phase $2\delta-\varphi_s$ shows that, in the Poincar\'e sphere of polarization, the signal and idler beams are mirror images of each other upon reflection on a plane resulting from rotation of the equatorial plane around the H/V axis by an angle $\delta$, as illustrated in Fig. \ref{fig:poincare} c).

\begin{figure}
\centering
   \includegraphics[width=\columnwidth]{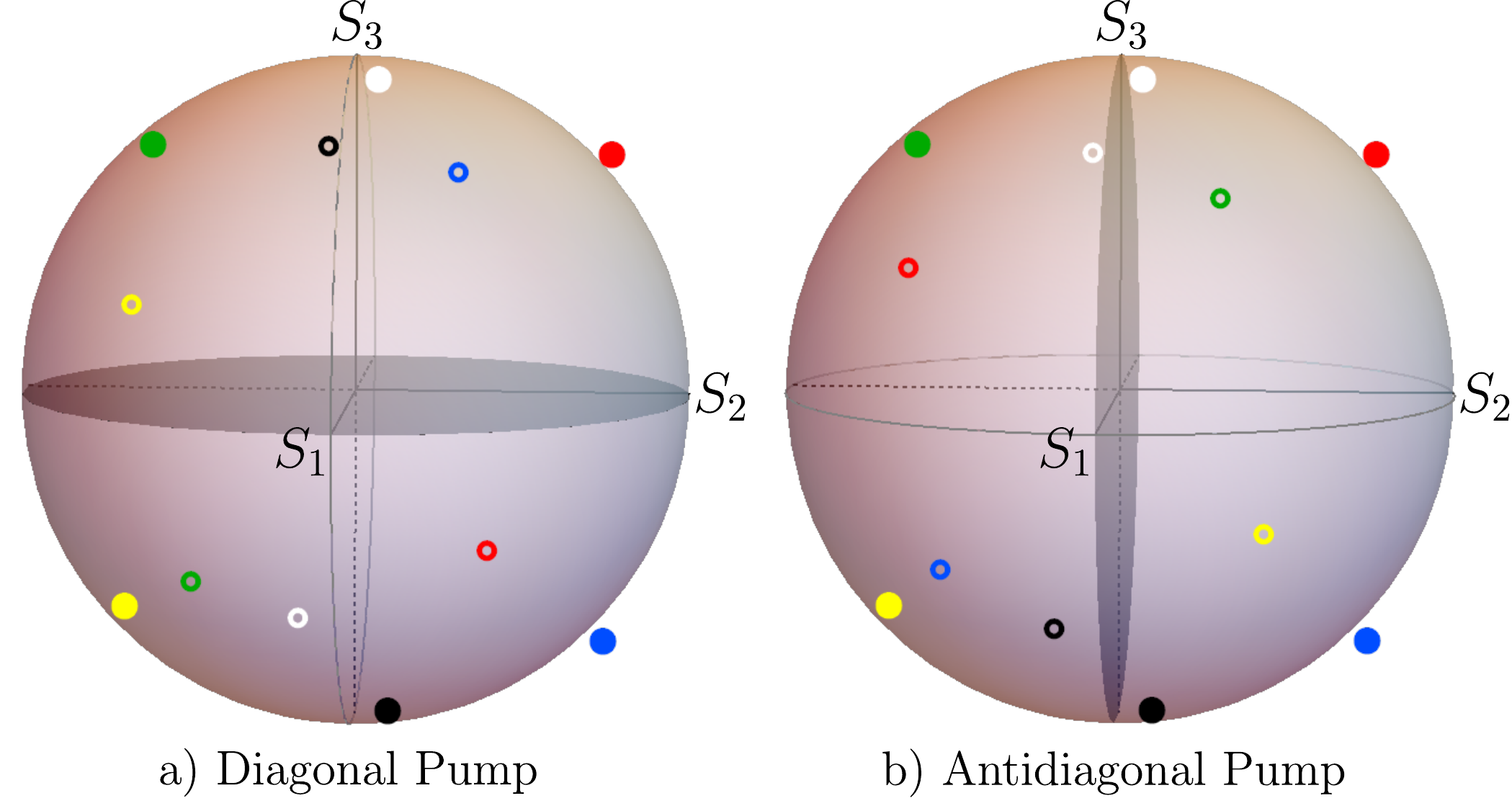}
   \caption{Poincar\'e spheres when pump beam polarization is linear a) diagonal and b) antidiagonal. Solid (open) circles correspond to the seed (idler). In both cases, the seed beam is prepared in six different polarization states: R (right-circular; white), L (left-circular; black), E$_1$, E$_2$, E$_3$ and E$_4$ (elliptical; red, blue, yellow and green, respectively), represented by filled circles. Blue and red discs seem outside the sphere surface due to uncertainties that are no represented in this picture. See Fig. \ref{fig:coordinates} for the error bars.}
   \label{fig:LREL}
\end{figure}

\paragraph{\bf{Experiment - isotropic polarization states}}
\label{sec:exp}
\begin{figure}
\centering
   \includegraphics[width=\columnwidth]{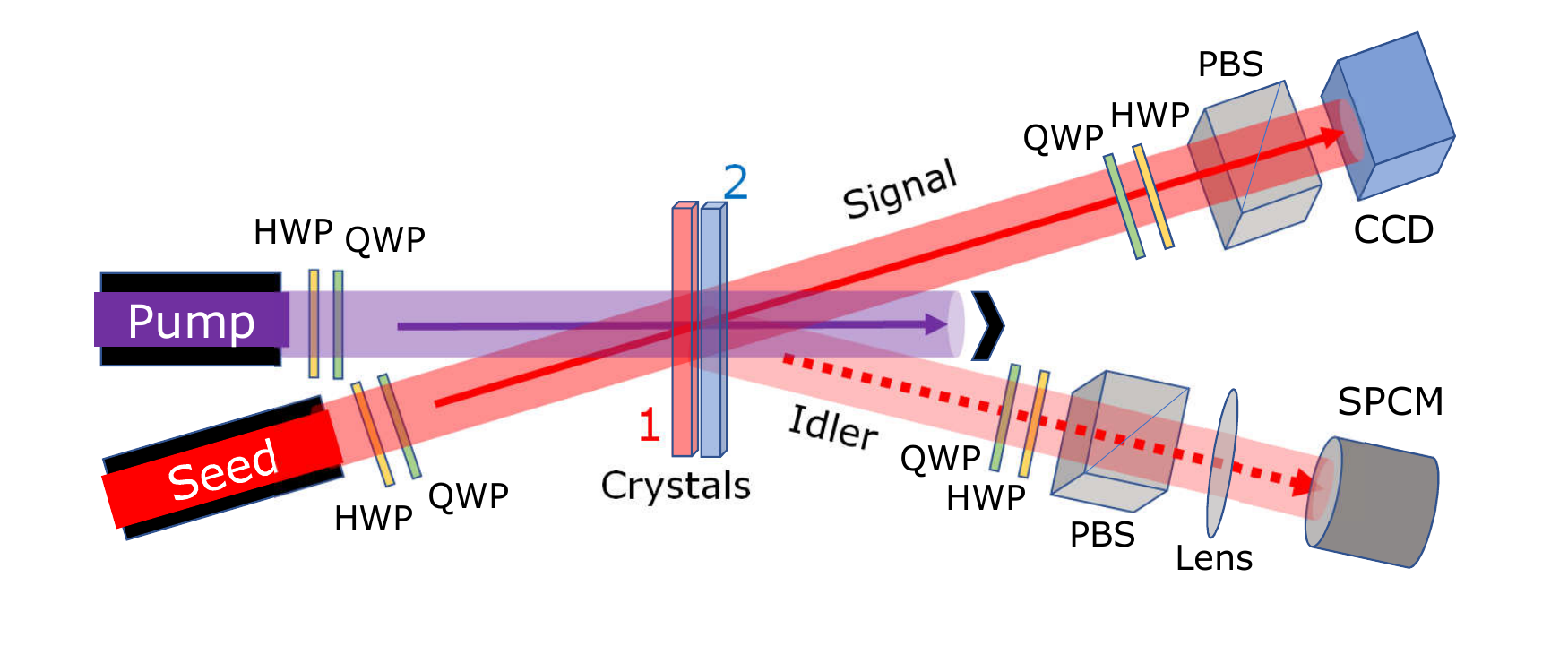}
   \caption{Experimental setup. A diode laser oscillating at the wavelength 405 nm is used to pump two identical type-I BBO ($\beta$ barium borate) crystals in sequence. The optical axis of the crystals are rotated by 90 degrees from each other, so that crystal 1 produces vertically polarized photon pairs and crystal 2 generates horizontally polarized photons.
This arrangement has been employed to generate polarization-entangled photon pairs \cite{Kwiat1999}, though here we will perform polarization-dependent stimulated emission. A seed diode laser at 780 nm is aligned to the desired signal direction and stimulates the conversion to the wavelengths 780 nm (signal-direct stimulation) and 840 nm (idler indirect stimulation), in non collinear phase-matching, so that signal and idler directions of propagation form an angle of about $4^\circ$.
The polarization states of the pump and seed beams are prepared using a half-wave plate (HWP) and a quarter-wave plate (QWP) for each beam. This allows us to pump and seed the parametric down-conversion process with arbitrary and controlled polarization states. The polarization state of the idler field, which is indirectly stimulated, is analyzed by measuring its Stokes parameters. This measurement is realized with an adjustable polarization analyzer, consisting of a QWP, a HWP and a polarizing beam splitter (PBS).  The idler is detected with a single-photon counting module(SPCM), in front of which there is 10 nm bandwidth interference filter centered at 840 nm. There is also a polarization analysis scheme for the seed, which is detected with a CCD camera.}
   \label{fig:setup}
\end{figure}
We realized phase conjugation using the experimental setup sketched in Fig. \ref{fig:setup}.
{We prepared the seed beam in six different polarization states and measured the corresponding idler beam polarization states, for both $A$- and $D$-polarized pump beams. Full polarization tomography of the seed/signal and idler beams was performed and the Stokes parameters were extracted, and plotted on the Poincar\'e sphere. 
\par
Results for isotropic polarization states are illustrated using the Poincar\'e sphere, shown in Fig. \ref{fig:LREL} a) for a $D$-polarized pump beam. We observe that seed (signal) and idler polarization states are located at opposite hemispheres, corresponding to phase conjugation. The conjugate of a given polarization state on the sphere is obtained by  changing the sign of the latitude while keeping the longitude ($\theta \longrightarrow \pi-\theta$). Our results clearly illustrate this effect for both circular and elliptical states.
\par
For a $A$-polarized pump beam, however, polarization conjugation does not occur, in the usual sense. Rather, the idler is close to the mirror image of the seed through the vertical plane $S_1S_3$, as shown in Fig. \ref{fig:LREL} b). The ability to control phase conjugation is an interesting feature, in contrast to the conjugation of the spatial degrees of freedom, which always occurs.
}
\begin{figure}
\centering

   \includegraphics[width=\columnwidth]{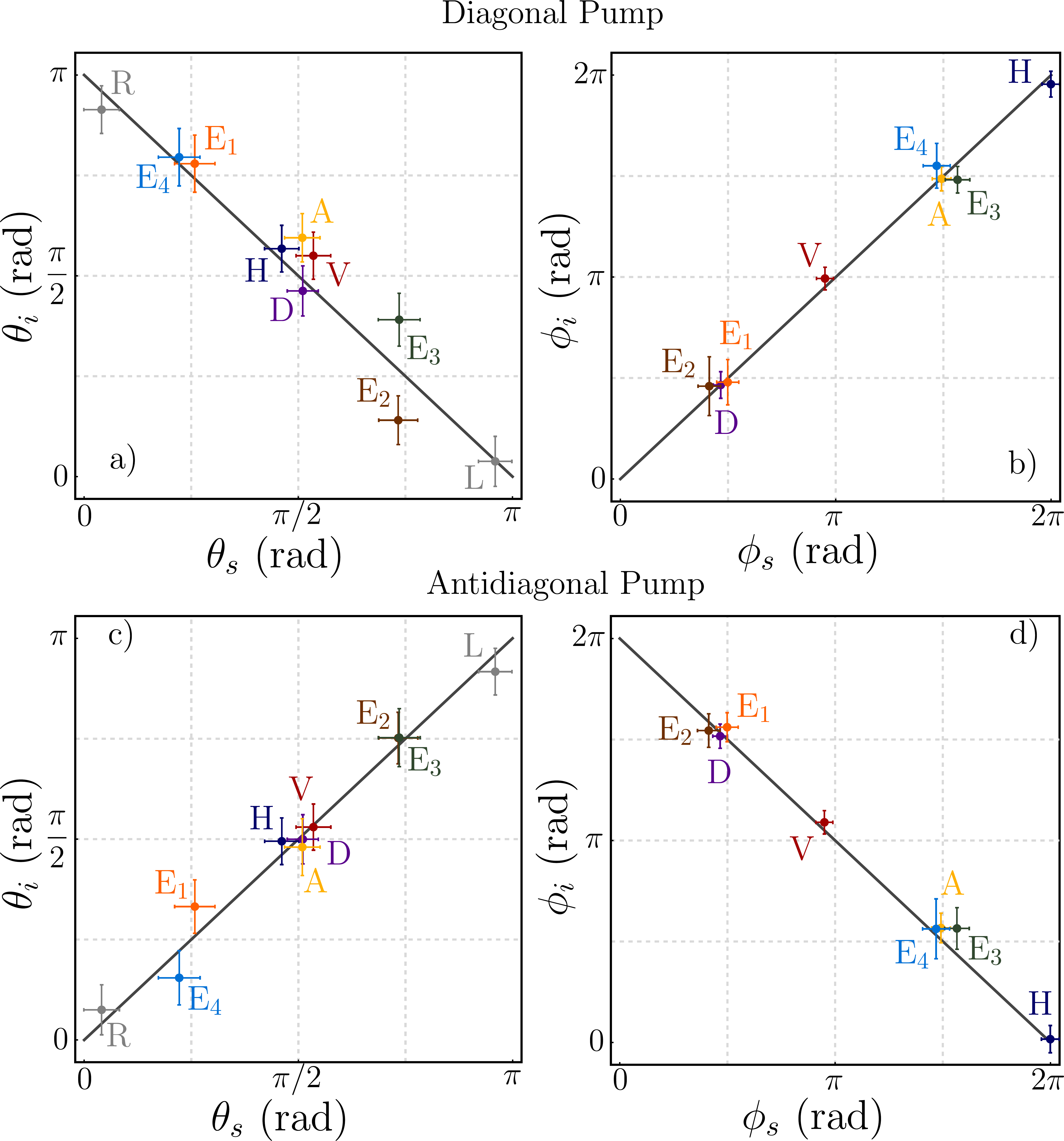}
   \caption{Spherical-coordinate angles: experimental data for idler's $\theta_i$ and $\phi_i$ versus signal's $\theta_s$ and $\phi_s$) for a)-b) diagonal pump and c)-d) antidiagonal pump. The solid straight lines represent the theoretical predictions.}
\label{fig:coordinates}
\end{figure}

For a more quantitative view, we plot idler versus seed/signal angular coordinates in the Poincar\'e sphere. The results (points with error bars)  are shown in  Fig. \ref{fig:coordinates} along with theory (solid line).  To compare polar ($\theta$) and azimuthal ($\phi$) coordinates, we assigned colors to each experimental point, with labels denoting polarization of the seed beam.

\begin{table}[t]
\centering
\resizebox{\linewidth}{!}{%
\begin{tabular}{cccccccccccl}
\multicolumn{12}{c}{\textbf{Diagonal Pump}} \\ 
\cline{1-11}
\multicolumn{1}{|c|}{Seed}     & \multicolumn{1}{c|}{R}        & \multicolumn{1}{c|}{E1}       & \multicolumn{1}{c|}{H}        & \multicolumn{1}{c|}{D}        & \multicolumn{1}{c|}{E2}       & \multicolumn{1}{c|}{L}        & \multicolumn{1}{c|}{E3}       & \multicolumn{1}{c|}{V}        & \multicolumn{1}{c|}{A}        & \multicolumn{1}{c|}{E4}       &                               \\ \hline
\multicolumn{1}{|c|}{Idler (theory)}   & \multicolumn{1}{c|}{L}        & \multicolumn{1}{c|}{E2}       & \multicolumn{1}{c|}{H}        & \multicolumn{1}{c|}{D}        & \multicolumn{1}{c|}{E1}       & \multicolumn{1}{c|}{R}        & \multicolumn{1}{c|}{E4}       & \multicolumn{1}{c|}{V}        & \multicolumn{1}{c|}{A}        & \multicolumn{1}{c|}{E3}       & \multicolumn{1}{l|}{Mean}     \\ \hline
\multicolumn{1}{|c|}{Fidelity (\%)} & \multicolumn{1}{c|}{86.4}  & \multicolumn{1}{c|}{81.3} & \multicolumn{1}{c|}{92.7} & \multicolumn{1}{c|}{84.1} & \multicolumn{1}{c|}{84.0} & \multicolumn{1}{c|}{86.6} & \multicolumn{1}{c|}{83.0} & \multicolumn{1}{c|}{93.7} & \multicolumn{1}{c|}{87.6} & \multicolumn{1}{c|}{83.4} & \multicolumn{1}{l|}{86.3}  \\ \hline
\multicolumn{1}{l}{}           & \multicolumn{1}{l}{}          & \multicolumn{1}{l}{}          & \multicolumn{1}{l}{}          & \multicolumn{1}{l}{}          & \multicolumn{1}{l}{}          & \multicolumn{1}{l}{}          & \multicolumn{1}{l}{}          & \multicolumn{1}{l}{}          & \multicolumn{1}{l}{}          & \multicolumn{1}{l}{}          &                               \\

\multicolumn{12}{c}{\textbf{Antidiagonal Pump}} \\ 
\cline{1-11}
\multicolumn{1}{|c|}{Seed}     & \multicolumn{1}{c|}{R}        & \multicolumn{1}{c|}{E1}       & \multicolumn{1}{c|}{H}        & \multicolumn{1}{c|}{D}        & \multicolumn{1}{c|}{E2}       & \multicolumn{1}{c|}{L}        & \multicolumn{1}{c|}{E3}       & \multicolumn{1}{c|}{V}        & \multicolumn{1}{c|}{A}        & \multicolumn{1}{c|}{E4}       & \multicolumn{1}{c}{}          \\ \hline
\multicolumn{1}{|c|}{Idler (theory)}   & \multicolumn{1}{c|}{R}        & \multicolumn{1}{c|}{E4}       & \multicolumn{1}{c|}{H}        & \multicolumn{1}{c|}{A}        & \multicolumn{1}{c|}{E3}       & \multicolumn{1}{c|}{L}        & \multicolumn{1}{c|}{E2}       & \multicolumn{1}{c|}{V}        & \multicolumn{1}{c|}{D}        & \multicolumn{1}{c|}{E1}       & \multicolumn{1}{c|}{Mean}     \\ \hline
\multicolumn{1}{|c|}{Fidelity (\%)} & \multicolumn{1}{c|}{85.4} & \multicolumn{1}{c|}{85.1}  & \multicolumn{1}{c|}{93.4} & \multicolumn{1}{c|}{87.6} & \multicolumn{1}{c|}{88.5}  & \multicolumn{1}{c|}{87.2} & \multicolumn{1}{c|}{81.6} & \multicolumn{1}{c|}{91.9} & \multicolumn{1}{c|}{80.3}  & \multicolumn{1}{c|}{80.9} & \multicolumn{1}{c|}{86.2} \\ \hline
\end{tabular}%
}
\caption{Fidelities between theoretical and measured idler polarization states pumping with diagonally and antidiagonally polarized beams for several seed preparations.}
\label{tab:table-fid}
\end{table}
 In addition, in Table \ref{tab:table-fid}, we list the fidelity \cite{jozsa1994fidelity,jerrard1982modern} of the measured idler's polarization state with respect to the theoretical prediction. All calculated fidelities lie between $80\%$ and $94\%$ with average of $86.3\%$ for diagonal pump and $86.2\%$ for antidiagonal pump. 
 Our experimental results clearly demonstrate phase conjugation of isotropic polarization states. Discrepancies with theory come mainly from depolarization of the measured light, which we attribute to scattering on the crystals surfaces as well as spatial walk-off, which could be reduced by using thinner crystals.

\paragraph{\bf{Experiment - anisotropic polarization states}}
\label{sec:anisotropic}

\begin{figure}
    \centering
    \includegraphics[width=0.75\columnwidth]{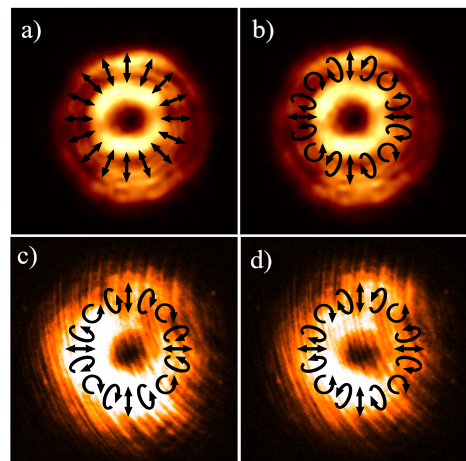}
      \caption{Vector vortex beams. Images were obtained experimentally. Arrows correspond to theoretical polarization directions. The radial vector beam a) is first transformed to the anisotropic beam b), used as the seed beam. The resulting idler phase conjugated c) and not conjugated d)}. 
    \label{fig:vortex}
\end{figure}

\begin{figure}
    \centering
    \includegraphics[width=0.75\columnwidth]{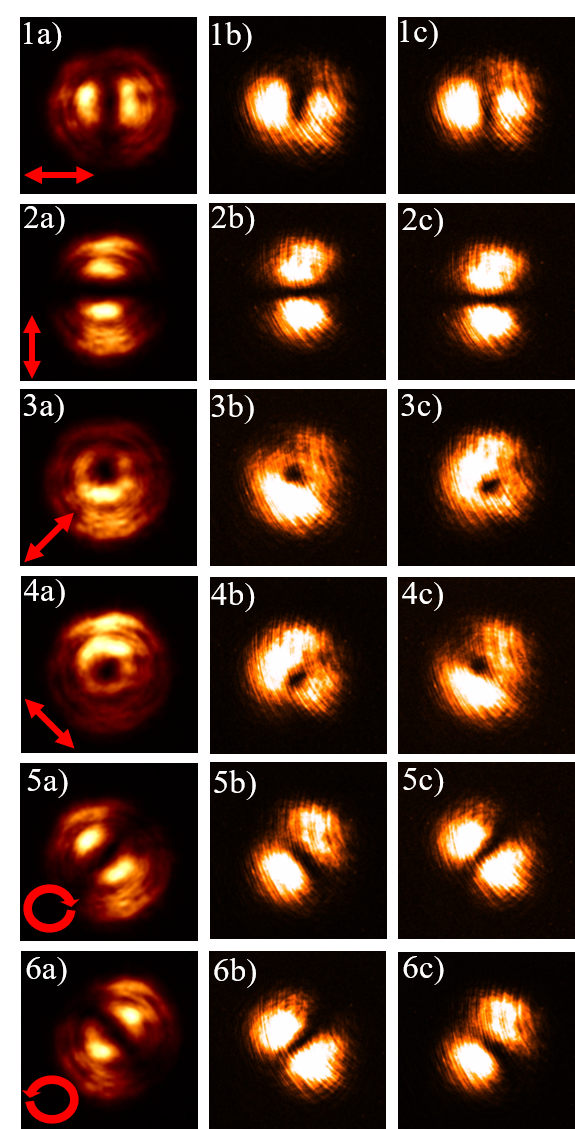} 
    \caption{Measurement results showing phase conjugation.  Polarization projections onto H/V, A/D, and R/L bases. 1a)--6a) seed beam, 1b)--6b) idler beam conjugated, 1c)--6c) idler beam not conjugated. } 
    \label{fig:conju}
\end{figure}

We now show that this setup realizes phase conjugation of vector beams.  A vortex half-wave plate (Thorlabs WPV10L-780) is placed in the path of the seed beam in Fig. \ref{fig:setup}, which produces a radial vector beam containing only linear polarization, as shown in Fig. \ref{fig:vortex}a). The profile was measured and the arrows indicate the polarization states in the profile. To better illustrate phase conjugation of the vector beam, we apply a  QWP to the radial beam so that a vortex beam containing both linear and circular polarization states are generated, as indicated in Fig. \ref{fig:vortex}b). This beam is used as the seed in StimPDC.   

Figs. \ref{fig:vortex}c) and d) show images of the doughnut-shaped intensity profile of the idler beam obtained in StimPDC when the pump beam is $D$- and $A$- polarized respectively. A signature of phase conjugation is the inversion of sense of rotation of the circularly polarized components (recall that both beams propagate forward).

The intensity profiles alone are not sufficient to show that phase conjugation occurs.  To do so, we perform polarization measurements on the seed and idler beams, and examine the images of the resulting beams using a CCD camera.  Fig. \ref{fig:conju} shows images of the seed (column a) and idler (column b for D-polarized pump and column c for A-polarized pump) beams, upon projections on the linear polarization bases H/V (rows 1 and 2)  and A/D (rows 3 and 4) and circular polarization basis R/L (rows 5 and 6).  Phase conjugation of the vector beam can be observed more clearly by comparing the images from projection in the circular polarization basis. Figs. \ref{fig:conju} 5b) and 6b) present a diagonal(anti-diagonal) Hermite-Gaussian shape contrasting with the anti-diagonal(diagonal) shape of the seed in Figs. \ref{fig:conju} 5a) and 6a) respectively, showing phase conjugation, while Figs. \ref{fig:conju} 5c) and 6c) show anti-diagonal(diagonal) Hermite-Gaussian shape, indicating no phase conjugation (again, recall that all beams propagate forward). At the same time, there is no difference between conjugation and no-conjugation, when the projections onto linear polarization bases are performed. 

\paragraph{\bf{Conclusion}}
\label{sec:conc}
Optical phase conjugation has concerned the preparation of a light beam that is the time reversal of another one in terms of its angular spectrum. Similarly, vector beam phase conjugation denotes time reversal that also includes the polarization degree of freedom.   We demonstrate theoretically and experimentally the process of vector beam phase conjugation using three-wave mixing in a two-crystal source.  The scheme is fast and can be conveniently controlled by manipulating the pump beam, and thus can be used in applications in which real-time corrections must be made to vector beams propagating in anisotropic and/or birefringent media.  We expect our results to broaden the range of possibilities for the use of vector beams in real world applications. 

\begin{acknowledgments}
The authors would like to thank the Brazilian Agencies CNPq, FAPESC, FAPERJ, FAPEG and the Brazilian National Institute of Science and Technology of Quantum Information (INCT/IQ). This study was funded in part by the Coordena\c{c}\~{a}o de Aperfei\c{c}oamento de Pessoal de N\'{i}vel Superior - Brasil (CAPES) - Finance Code 001.
\end{acknowledgments}

\bibliographystyle{apsrev}
\bibliography{StimPDCpol}

\begin{thebibliography}{46}
\expandafter\ifx\csname natexlab\endcsname\relax\def\natexlab#1{#1}\fi
\expandafter\ifx\csname bibnamefont\endcsname\relax
  \def\bibnamefont#1{#1}\fi
\expandafter\ifx\csname bibfnamefont\endcsname\relax
  \def\bibfnamefont#1{#1}\fi
\expandafter\ifx\csname citenamefont\endcsname\relax
  \def\citenamefont#1{#1}\fi
\expandafter\ifx\csname url\endcsname\relax
  \def\url#1{\texttt{#1}}\fi
\expandafter\ifx\csname urlprefix\endcsname\relax\def\urlprefix{URL }\fi
\providecommand{\bibinfo}[2]{#2}
\providecommand{\eprint}[2][]{\url{#2}}

\bibitem[{\citenamefont{D'Ambrosio et~al.}(2012)\citenamefont{D'Ambrosio,
  Nagali, Walborn, Aolita, Slussarenko, Marrucci, and Sciarrino}}]{dambrosio12}
\bibinfo{author}{\bibfnamefont{V.}~\bibnamefont{D'Ambrosio}},
  \bibinfo{author}{\bibfnamefont{E.}~\bibnamefont{Nagali}},
  \bibinfo{author}{\bibfnamefont{S.~P.} \bibnamefont{Walborn}},
  \bibinfo{author}{\bibfnamefont{L.}~\bibnamefont{Aolita}},
  \bibinfo{author}{\bibfnamefont{S.}~\bibnamefont{Slussarenko}},
  \bibinfo{author}{\bibfnamefont{L.}~\bibnamefont{Marrucci}}, \bibnamefont{and}
  \bibinfo{author}{\bibfnamefont{F.}~\bibnamefont{Sciarrino}},
  \bibinfo{journal}{Nature Communications} \textbf{\bibinfo{volume}{3}},
  \bibinfo{pages}{961} (\bibinfo{year}{2012}).

\bibitem[{\citenamefont{Zhao and Wang}(2015)}]{Zhao15}
\bibinfo{author}{\bibfnamefont{Y.}~\bibnamefont{Zhao}} \bibnamefont{and}
  \bibinfo{author}{\bibfnamefont{J.}~\bibnamefont{Wang}},
  \bibinfo{journal}{Opt. Lett.} \textbf{\bibinfo{volume}{40}},
  \bibinfo{pages}{4843} (\bibinfo{year}{2015}).

\bibitem[{\citenamefont{as et~al.}(2015)\citenamefont{as, D'Ambrosio,
  Taballione, Bisesto, Slussarenko, Aolita, Marrucci, Walborn, and
  Sciarrino}}]{Farias15}
\bibinfo{author}{\bibfnamefont{O.~J.~F.} \bibnamefont{as}},
  \bibinfo{author}{\bibfnamefont{V.}~\bibnamefont{D'Ambrosio}},
  \bibinfo{author}{\bibfnamefont{C.}~\bibnamefont{Taballione}},
  \bibinfo{author}{\bibfnamefont{F.}~\bibnamefont{Bisesto}},
  \bibinfo{author}{\bibfnamefont{S.}~\bibnamefont{Slussarenko}},
  \bibinfo{author}{\bibfnamefont{L.}~\bibnamefont{Aolita}},
  \bibinfo{author}{\bibfnamefont{L.}~\bibnamefont{Marrucci}},
  \bibinfo{author}{\bibfnamefont{S.~P.} \bibnamefont{Walborn}},
  \bibnamefont{and}
  \bibinfo{author}{\bibfnamefont{F.}~\bibnamefont{Sciarrino}},
  \bibinfo{journal}{Sci. Rep.} \textbf{\bibinfo{volume}{5}},
  \bibinfo{pages}{8424} (\bibinfo{year}{2015}).

\bibitem[{\citenamefont{Milione
  et~al.}(2015{\natexlab{a}})\citenamefont{Milione, Nguyen, Leach, Nolan, and
  Alfano}}]{Milione15}
\bibinfo{author}{\bibfnamefont{G.}~\bibnamefont{Milione}},
  \bibinfo{author}{\bibfnamefont{T.~A.} \bibnamefont{Nguyen}},
  \bibinfo{author}{\bibfnamefont{J.}~\bibnamefont{Leach}},
  \bibinfo{author}{\bibfnamefont{D.~A.} \bibnamefont{Nolan}}, \bibnamefont{and}
  \bibinfo{author}{\bibfnamefont{R.~R.} \bibnamefont{Alfano}},
  \bibinfo{journal}{Opt. Lett.} \textbf{\bibinfo{volume}{40}},
  \bibinfo{pages}{4887} (\bibinfo{year}{2015}{\natexlab{a}}).

\bibitem[{\citenamefont{Milione
  et~al.}(2015{\natexlab{b}})\citenamefont{Milione, Lavery, Huang, Ren, Xie,
  Nguyen, Karimi, Marrucci, Nolan, Alfano et~al.}}]{Milione15A}
\bibinfo{author}{\bibfnamefont{G.}~\bibnamefont{Milione}},
  \bibinfo{author}{\bibfnamefont{M.~P.~J.} \bibnamefont{Lavery}},
  \bibinfo{author}{\bibfnamefont{H.}~\bibnamefont{Huang}},
  \bibinfo{author}{\bibfnamefont{Y.}~\bibnamefont{Ren}},
  \bibinfo{author}{\bibfnamefont{G.}~\bibnamefont{Xie}},
  \bibinfo{author}{\bibfnamefont{T.~A.} \bibnamefont{Nguyen}},
  \bibinfo{author}{\bibfnamefont{E.}~\bibnamefont{Karimi}},
  \bibinfo{author}{\bibfnamefont{L.}~\bibnamefont{Marrucci}},
  \bibinfo{author}{\bibfnamefont{D.~A.} \bibnamefont{Nolan}},
  \bibinfo{author}{\bibfnamefont{R.~R.} \bibnamefont{Alfano}},
  \bibnamefont{et~al.}, \bibinfo{journal}{Opt. Lett.}
  \textbf{\bibinfo{volume}{40}}, \bibinfo{pages}{1980}
  (\bibinfo{year}{2015}{\natexlab{b}}).

\bibitem[{\citenamefont{Zhang et~al.}(2016)\citenamefont{Zhang, Li, Li, Feng,
  and Li}}]{zhang16}
\bibinfo{author}{\bibfnamefont{J.}~\bibnamefont{Zhang}},
  \bibinfo{author}{\bibfnamefont{F.}~\bibnamefont{Li}},
  \bibinfo{author}{\bibfnamefont{J.}~\bibnamefont{Li}},
  \bibinfo{author}{\bibfnamefont{Y.}~\bibnamefont{Feng}}, \bibnamefont{and}
  \bibinfo{author}{\bibfnamefont{Z.}~\bibnamefont{Li}}, \bibinfo{journal}{IEEE
  Photon. J.} \textbf{\bibinfo{volume}{8}}, \bibinfo{pages}{7907008}
  (\bibinfo{year}{2016}).

\bibitem[{\citenamefont{Li et~al.}(2016)\citenamefont{Li, Wang, and
  Zhang}}]{Li16}
\bibinfo{author}{\bibfnamefont{P.}~\bibnamefont{Li}},
  \bibinfo{author}{\bibfnamefont{B.}~\bibnamefont{Wang}}, \bibnamefont{and}
  \bibinfo{author}{\bibfnamefont{X.}~\bibnamefont{Zhang}},
  \bibinfo{journal}{Opt. Express} \textbf{\bibinfo{volume}{24}},
  \bibinfo{pages}{15143} (\bibinfo{year}{2016}).

\bibitem[{\citenamefont{Ndagano et~al.}(2018)\citenamefont{Ndagano, Nape, Cos,
  Rosales-Guzman, and Forbes}}]{Ndagano18}
\bibinfo{author}{\bibfnamefont{B.}~\bibnamefont{Ndagano}},
  \bibinfo{author}{\bibfnamefont{I.}~\bibnamefont{Nape}},
  \bibinfo{author}{\bibfnamefont{M.~A.} \bibnamefont{Cos}},
  \bibinfo{author}{\bibfnamefont{C.}~\bibnamefont{Rosales-Guzman}},
  \bibnamefont{and} \bibinfo{author}{\bibfnamefont{A.}~\bibnamefont{Forbes}},
  \bibinfo{journal}{J. Lightwave Tech.} \textbf{\bibinfo{volume}{36}},
  \bibinfo{pages}{292} (\bibinfo{year}{2018}).

\bibitem[{\citenamefont{Gregg et~al.}(2015)\citenamefont{Gregg, Mirhosseini,
  Rubano, Marrucci, Karimi, Boyd, and Ramachandran}}]{Gregg15}
\bibinfo{author}{\bibfnamefont{P.}~\bibnamefont{Gregg}},
  \bibinfo{author}{\bibfnamefont{M.}~\bibnamefont{Mirhosseini}},
  \bibinfo{author}{\bibfnamefont{A.}~\bibnamefont{Rubano}},
  \bibinfo{author}{\bibfnamefont{L.}~\bibnamefont{Marrucci}},
  \bibinfo{author}{\bibfnamefont{E.}~\bibnamefont{Karimi}},
  \bibinfo{author}{\bibfnamefont{R.~W.} \bibnamefont{Boyd}}, \bibnamefont{and}
  \bibinfo{author}{\bibfnamefont{S.}~\bibnamefont{Ramachandran}},
  \bibinfo{journal}{Opt. Lett.} \textbf{\bibinfo{volume}{40}},
  \bibinfo{pages}{1729} (\bibinfo{year}{2015}).

\bibitem[{\citenamefont{Ndagano et~al.}(2015)\citenamefont{Ndagano,
  Br\"{u}ning, McLaren, Duparr\'{e}, and Forbes}}]{Ndagano15}
\bibinfo{author}{\bibfnamefont{B.}~\bibnamefont{Ndagano}},
  \bibinfo{author}{\bibfnamefont{R.}~\bibnamefont{Br\"{u}ning}},
  \bibinfo{author}{\bibfnamefont{M.}~\bibnamefont{McLaren}},
  \bibinfo{author}{\bibfnamefont{M.}~\bibnamefont{Duparr\'{e}}},
  \bibnamefont{and} \bibinfo{author}{\bibfnamefont{A.}~\bibnamefont{Forbes}},
  \bibinfo{journal}{Opt. Express} \textbf{\bibinfo{volume}{23}},
  \bibinfo{pages}{17330} (\bibinfo{year}{2015}).

\bibitem[{\citenamefont{Biss et~al.}(2006)\citenamefont{Biss, Youngworth, and
  Brown}}]{Biss06}
\bibinfo{author}{\bibfnamefont{D.~P.} \bibnamefont{Biss}},
  \bibinfo{author}{\bibfnamefont{K.~S.} \bibnamefont{Youngworth}},
  \bibnamefont{and} \bibinfo{author}{\bibfnamefont{T.~G.} \bibnamefont{Brown}},
  \bibinfo{journal}{Appl. Opt.} \textbf{\bibinfo{volume}{45}},
  \bibinfo{pages}{470} (\bibinfo{year}{2006}).

\bibitem[{\citenamefont{Yoshida et~al.}(2019)\citenamefont{Yoshida, Kozawa, and
  Sato}}]{Yoshida19}
\bibinfo{author}{\bibfnamefont{M.}~\bibnamefont{Yoshida}},
  \bibinfo{author}{\bibfnamefont{Y.}~\bibnamefont{Kozawa}}, \bibnamefont{and}
  \bibinfo{author}{\bibfnamefont{S.}~\bibnamefont{Sato}},
  \bibinfo{journal}{Opt. Lett.} \textbf{\bibinfo{volume}{44}},
  \bibinfo{pages}{883} (\bibinfo{year}{2019}).

\bibitem[{\citenamefont{D'Ambrosio et~al.}(2013)\citenamefont{D'Ambrosio,
  Spagnolo, Re, Slussarenko, Li, Kwek, Marrucci, Walborn, Aolita, and
  Sciarrino}}]{dambrosio13b}
\bibinfo{author}{\bibfnamefont{V.}~\bibnamefont{D'Ambrosio}},
  \bibinfo{author}{\bibfnamefont{N.}~\bibnamefont{Spagnolo}},
  \bibinfo{author}{\bibfnamefont{L.~D.} \bibnamefont{Re}},
  \bibinfo{author}{\bibfnamefont{S.}~\bibnamefont{Slussarenko}},
  \bibinfo{author}{\bibfnamefont{Y.}~\bibnamefont{Li}},
  \bibinfo{author}{\bibfnamefont{L.~C.} \bibnamefont{Kwek}},
  \bibinfo{author}{\bibfnamefont{L.}~\bibnamefont{Marrucci}},
  \bibinfo{author}{\bibfnamefont{S.~P.} \bibnamefont{Walborn}},
  \bibinfo{author}{\bibfnamefont{L.}~\bibnamefont{Aolita}}, \bibnamefont{and}
  \bibinfo{author}{\bibfnamefont{F.}~\bibnamefont{Sciarrino}},
  \bibinfo{journal}{Nature Communications} \textbf{\bibinfo{volume}{4}},
  \bibinfo{pages}{2432} (\bibinfo{year}{2013}).

\bibitem[{\citenamefont{Töppel et~al.}(2014)\citenamefont{Töppel, Aiello,
  Marquardt, Giacobino, and Leuchs}}]{Toppel2014}
\bibinfo{author}{\bibfnamefont{F.}~\bibnamefont{Töppel}},
  \bibinfo{author}{\bibfnamefont{A.}~\bibnamefont{Aiello}},
  \bibinfo{author}{\bibfnamefont{C.}~\bibnamefont{Marquardt}},
  \bibinfo{author}{\bibfnamefont{E.}~\bibnamefont{Giacobino}},
  \bibnamefont{and} \bibinfo{author}{\bibfnamefont{G.}~\bibnamefont{Leuchs}},
  \bibinfo{journal}{New Journal of Physics} \textbf{\bibinfo{volume}{16}},
  \bibinfo{pages}{073019} (\bibinfo{year}{2014}).

\bibitem[{\citenamefont{Berg-Johansen et~al.}(2015)\citenamefont{Berg-Johansen,
  T\"{o}ppel, Stiller, Banzer, Ornigotti, Giacobino, Leuchs, Aiello, and
  Marquardt}}]{Berg-Johansen15}
\bibinfo{author}{\bibfnamefont{S.}~\bibnamefont{Berg-Johansen}},
  \bibinfo{author}{\bibfnamefont{F.}~\bibnamefont{T\"{o}ppel}},
  \bibinfo{author}{\bibfnamefont{B.}~\bibnamefont{Stiller}},
  \bibinfo{author}{\bibfnamefont{P.}~\bibnamefont{Banzer}},
  \bibinfo{author}{\bibfnamefont{M.}~\bibnamefont{Ornigotti}},
  \bibinfo{author}{\bibfnamefont{E.}~\bibnamefont{Giacobino}},
  \bibinfo{author}{\bibfnamefont{G.}~\bibnamefont{Leuchs}},
  \bibinfo{author}{\bibfnamefont{A.}~\bibnamefont{Aiello}}, \bibnamefont{and}
  \bibinfo{author}{\bibfnamefont{C.}~\bibnamefont{Marquardt}},
  \bibinfo{journal}{Optica} \textbf{\bibinfo{volume}{2}}, \bibinfo{pages}{864}
  (\bibinfo{year}{2015}).

\bibitem[{\citenamefont{Zhan}(2009)}]{Zhan09}
\bibinfo{author}{\bibfnamefont{Q.}~\bibnamefont{Zhan}}, \bibinfo{journal}{Adv.
  Opt. Photon.} \textbf{\bibinfo{volume}{1}}, \bibinfo{pages}{1}
  (\bibinfo{year}{2009}).

\bibitem[{\citenamefont{Chen et~al.}(2013)\citenamefont{Chen, Agarwal,
  Sheppard, and Chen}}]{Chen13}
\bibinfo{author}{\bibfnamefont{R.}~\bibnamefont{Chen}},
  \bibinfo{author}{\bibfnamefont{K.}~\bibnamefont{Agarwal}},
  \bibinfo{author}{\bibfnamefont{C.}~\bibnamefont{Sheppard}}, \bibnamefont{and}
  \bibinfo{author}{\bibfnamefont{X.}~\bibnamefont{Chen}},
  \bibinfo{journal}{Optics Letters} \textbf{\bibinfo{volume}{38}},
  \bibinfo{pages}{3111} (\bibinfo{year}{2013}).

\bibitem[{\citenamefont{Niziev and Nesterov}(1999)}]{Niziev1999}
\bibinfo{author}{\bibfnamefont{V.~G.} \bibnamefont{Niziev}} \bibnamefont{and}
  \bibinfo{author}{\bibfnamefont{A.~V.} \bibnamefont{Nesterov}},
  \bibinfo{journal}{Journal of Physics D: Applied Physics}
  \textbf{\bibinfo{volume}{32}}, \bibinfo{pages}{1455} (\bibinfo{year}{1999}).

\bibitem[{\citenamefont{Schultz et~al.}(2009)\citenamefont{Schultz, Stranick,
  and Levin}}]{Schultz09}
\bibinfo{author}{\bibfnamefont{Z.~D.} \bibnamefont{Schultz}},
  \bibinfo{author}{\bibfnamefont{S.~J.} \bibnamefont{Stranick}},
  \bibnamefont{and} \bibinfo{author}{\bibfnamefont{I.~W.} \bibnamefont{Levin}},
  \bibinfo{journal}{Anal Chem.} \textbf{\bibinfo{volume}{81}},
  \bibinfo{pages}{9657} (\bibinfo{year}{2009}).

\bibitem[{\citenamefont{Kazemi-Zanjani
  et~al.}(2013)\citenamefont{Kazemi-Zanjani, Vedraine, and
  Lagugn\'{e}-Labarthet}}]{Kazemi-Zanjani13}
\bibinfo{author}{\bibfnamefont{N.}~\bibnamefont{Kazemi-Zanjani}},
  \bibinfo{author}{\bibfnamefont{S.}~\bibnamefont{Vedraine}}, \bibnamefont{and}
  \bibinfo{author}{\bibfnamefont{F.}~\bibnamefont{Lagugn\'{e}-Labarthet}},
  \bibinfo{journal}{Opt. Express} \textbf{\bibinfo{volume}{21}},
  \bibinfo{pages}{25271} (\bibinfo{year}{2013}).

\bibitem[{\citenamefont{Lu et~al.}(2018)\citenamefont{Lu, Huang, Lei, Su, Wang,
  Liu, Zhang, Wang, and Mei}}]{Lu18}
\bibinfo{author}{\bibfnamefont{F.}~\bibnamefont{Lu}},
  \bibinfo{author}{\bibfnamefont{T.-X.} \bibnamefont{Huang}},
  \bibinfo{author}{\bibfnamefont{H.}~\bibnamefont{Lei}},
  \bibinfo{author}{\bibfnamefont{H.}~\bibnamefont{Su}},
  \bibinfo{author}{\bibfnamefont{H.}~\bibnamefont{Wang}},
  \bibinfo{author}{\bibfnamefont{M.}~\bibnamefont{Liu}},
  \bibinfo{author}{\bibfnamefont{W.}~\bibnamefont{Zhang}},
  \bibinfo{author}{\bibfnamefont{X.}~\bibnamefont{Wang}}, \bibnamefont{and}
  \bibinfo{author}{\bibfnamefont{T.}~\bibnamefont{Mei}},
  \bibinfo{journal}{Sensors} \textbf{\bibinfo{volume}{18}},
  \bibinfo{pages}{3841} (\bibinfo{year}{2018}).

\bibitem[{\citenamefont{Fisher}(2012)}]{Fisher2012}
\bibinfo{author}{\bibfnamefont{R.~A.} \bibnamefont{Fisher}},
  \emph{\bibinfo{title}{Optical phase conjugation}}
  (\bibinfo{publisher}{Academic Press}, \bibinfo{year}{2012}).

\bibitem[{\citenamefont{MacDonald et~al.}(1988)\citenamefont{MacDonald,
  Tompkin, and Boyd}}]{MacDonald1988}
\bibinfo{author}{\bibfnamefont{K.~R.} \bibnamefont{MacDonald}},
  \bibinfo{author}{\bibfnamefont{W.~R.} \bibnamefont{Tompkin}},
  \bibnamefont{and} \bibinfo{author}{\bibfnamefont{R.~W.} \bibnamefont{Boyd}},
  \bibinfo{journal}{Optics letters} \textbf{\bibinfo{volume}{13}},
  \bibinfo{pages}{485} (\bibinfo{year}{1988}).

\bibitem[{\citenamefont{Boyd et~al.}(1989)\citenamefont{Boyd, MacDonald, and
  Malcuit}}]{Boyd1989}
\bibinfo{author}{\bibfnamefont{R.~W.} \bibnamefont{Boyd}},
  \bibinfo{author}{\bibfnamefont{K.~R.} \bibnamefont{MacDonald}},
  \bibnamefont{and} \bibinfo{author}{\bibfnamefont{M.~S.}
  \bibnamefont{Malcuit}}, in \emph{\bibinfo{booktitle}{Laser Wavefront
  Control}} (\bibinfo{organization}{International Society for Optics and
  Photonics}, \bibinfo{year}{1989}), vol. \bibinfo{volume}{1000}, pp.
  \bibinfo{pages}{69--81}.

\bibitem[{\citenamefont{Zel'Dovich et~al.}(1995)\citenamefont{Zel'Dovich,
  Popovichev, Ragul'Skii, and Faizullov}}]{Zel1995}
\bibinfo{author}{\bibfnamefont{B.~Y.} \bibnamefont{Zel'Dovich}},
  \bibinfo{author}{\bibfnamefont{V.}~\bibnamefont{Popovichev}},
  \bibinfo{author}{\bibfnamefont{V.}~\bibnamefont{Ragul'Skii}},
  \bibnamefont{and}
  \bibinfo{author}{\bibfnamefont{F.}~\bibnamefont{Faizullov}}, in
  \emph{\bibinfo{booktitle}{Landmark Papers on Photorefractive Nonlinear
  Optics}} (\bibinfo{publisher}{World Scientific}, \bibinfo{year}{1995}), pp.
  \bibinfo{pages}{303--306}.

\bibitem[{\citenamefont{Levenson}(1980)}]{Levenson1980}
\bibinfo{author}{\bibfnamefont{M.~D.} \bibnamefont{Levenson}},
  \bibinfo{journal}{Optics letters} \textbf{\bibinfo{volume}{5}},
  \bibinfo{pages}{182} (\bibinfo{year}{1980}).

\bibitem[{\citenamefont{McFarlane and Steel}(1983)}]{McFarlane1983}
\bibinfo{author}{\bibfnamefont{R.}~\bibnamefont{McFarlane}} \bibnamefont{and}
  \bibinfo{author}{\bibfnamefont{D.}~\bibnamefont{Steel}},
  \bibinfo{journal}{Optics letters} \textbf{\bibinfo{volume}{8}},
  \bibinfo{pages}{208} (\bibinfo{year}{1983}).

\bibitem[{\citenamefont{Gower}(1982)}]{Gower1982}
\bibinfo{author}{\bibfnamefont{M.}~\bibnamefont{Gower}},
  \bibinfo{journal}{Optics letters} \textbf{\bibinfo{volume}{7}},
  \bibinfo{pages}{423} (\bibinfo{year}{1982}).

\bibitem[{\citenamefont{Mosk et~al.}(2012)\citenamefont{Mosk, Lagendijk,
  Lerosey, and Fink}}]{Mosk2012}
\bibinfo{author}{\bibfnamefont{A.~P.} \bibnamefont{Mosk}},
  \bibinfo{author}{\bibfnamefont{A.}~\bibnamefont{Lagendijk}},
  \bibinfo{author}{\bibfnamefont{G.}~\bibnamefont{Lerosey}}, \bibnamefont{and}
  \bibinfo{author}{\bibfnamefont{M.}~\bibnamefont{Fink}},
  \bibinfo{journal}{Nature photonics} \textbf{\bibinfo{volume}{6}},
  \bibinfo{pages}{283} (\bibinfo{year}{2012}).

\bibitem[{\citenamefont{Yaqoob et~al.}(2008)\citenamefont{Yaqoob, Psaltis,
  Feld, and Yang}}]{Yaqoob2008}
\bibinfo{author}{\bibfnamefont{Z.}~\bibnamefont{Yaqoob}},
  \bibinfo{author}{\bibfnamefont{D.}~\bibnamefont{Psaltis}},
  \bibinfo{author}{\bibfnamefont{M.~S.} \bibnamefont{Feld}}, \bibnamefont{and}
  \bibinfo{author}{\bibfnamefont{C.}~\bibnamefont{Yang}},
  \bibinfo{journal}{Nature photonics} \textbf{\bibinfo{volume}{2}},
  \bibinfo{pages}{110} (\bibinfo{year}{2008}).

\bibitem[{\citenamefont{Set et~al.}(1998)\citenamefont{Set, Yamashita, Ibsen,
  Laming, Nesset, Kelly, and Gilbertas}}]{Set98}
\bibinfo{author}{\bibfnamefont{S.}~\bibnamefont{Set}},
  \bibinfo{author}{\bibfnamefont{S.}~\bibnamefont{Yamashita}},
  \bibinfo{author}{\bibfnamefont{M.}~\bibnamefont{Ibsen}},
  \bibinfo{author}{\bibfnamefont{R.}~\bibnamefont{Laming}},
  \bibinfo{author}{\bibfnamefont{D.}~\bibnamefont{Nesset}},
  \bibinfo{author}{\bibfnamefont{A.}~\bibnamefont{Kelly}}, \bibnamefont{and}
  \bibinfo{author}{\bibfnamefont{C.}~\bibnamefont{Gilbertas}},
  \bibinfo{journal}{Electronics Letters} \textbf{\bibinfo{volume}{34}},
  \bibinfo{pages}{1681} (\bibinfo{year}{1998}).

\bibitem[{\citenamefont{Yariv and Pepper}(1977)}]{Yariv1977}
\bibinfo{author}{\bibfnamefont{A.}~\bibnamefont{Yariv}} \bibnamefont{and}
  \bibinfo{author}{\bibfnamefont{D.~M.} \bibnamefont{Pepper}},
  \bibinfo{journal}{Optics Letters} \textbf{\bibinfo{volume}{1}},
  \bibinfo{pages}{16} (\bibinfo{year}{1977}).

\bibitem[{\citenamefont{Green et~al.}(1996)\citenamefont{Green, Udaiyan,
  Crofts, Kim, and Damzen}}]{Damzen1996}
\bibinfo{author}{\bibfnamefont{R.~M.} \bibnamefont{Green}},
  \bibinfo{author}{\bibfnamefont{D.}~\bibnamefont{Udaiyan}},
  \bibinfo{author}{\bibfnamefont{G.}~\bibnamefont{Crofts}},
  \bibinfo{author}{\bibfnamefont{D.}~\bibnamefont{Kim}}, \bibnamefont{and}
  \bibinfo{author}{\bibfnamefont{M.}~\bibnamefont{Damzen}},
  \bibinfo{journal}{Physical Review Letters} \textbf{\bibinfo{volume}{77}},
  \bibinfo{pages}{3533} (\bibinfo{year}{1996}).

\bibitem[{\citenamefont{G{\"u}nter}(1985)}]{Gunter2007}
\bibinfo{author}{\bibfnamefont{P.}~\bibnamefont{G{\"u}nter}}, in
  \emph{\bibinfo{booktitle}{Festk{\"o}rperprobleme 25}}
  (\bibinfo{publisher}{Springer}, \bibinfo{year}{1985}), pp.
  \bibinfo{pages}{363--369}.

\bibitem[{\citenamefont{He et~al.}(2002)\citenamefont{He, Markowicz, Lin, and
  Prasad}}]{He2002}
\bibinfo{author}{\bibfnamefont{G.~S.} \bibnamefont{He}},
  \bibinfo{author}{\bibfnamefont{P.~P.} \bibnamefont{Markowicz}},
  \bibinfo{author}{\bibfnamefont{T.-C.} \bibnamefont{Lin}}, \bibnamefont{and}
  \bibinfo{author}{\bibfnamefont{P.~N.} \bibnamefont{Prasad}},
  \bibinfo{journal}{Nature} \textbf{\bibinfo{volume}{415}},
  \bibinfo{pages}{767} (\bibinfo{year}{2002}).

\bibitem[{\citenamefont{Qian et~al.}(2014{\natexlab{a}})\citenamefont{Qian, Li,
  Kong, and Tu}}]{Qian14_OL1}
\bibinfo{author}{\bibfnamefont{S.-X.} \bibnamefont{Qian}},
  \bibinfo{author}{\bibfnamefont{Y.}~\bibnamefont{Li}},
  \bibinfo{author}{\bibfnamefont{L.-J.} \bibnamefont{Kong}}, \bibnamefont{and}
  \bibinfo{author}{\bibfnamefont{C.}~\bibnamefont{Tu}},
  \bibinfo{journal}{Optics Letters} \textbf{\bibinfo{volume}{39}},
  \bibinfo{pages}{4907} (\bibinfo{year}{2014}{\natexlab{a}}).

\bibitem[{\citenamefont{Qian et~al.}(2014{\natexlab{b}})\citenamefont{Qian, Li,
  Kong, and Tu}}]{Qian14_OL2}
\bibinfo{author}{\bibfnamefont{S.-X.} \bibnamefont{Qian}},
  \bibinfo{author}{\bibfnamefont{Y.}~\bibnamefont{Li}},
  \bibinfo{author}{\bibfnamefont{L.-J.} \bibnamefont{Kong}}, \bibnamefont{and}
  \bibinfo{author}{\bibfnamefont{C.}~\bibnamefont{Tu}},
  \bibinfo{journal}{Optics Letters} \textbf{\bibinfo{volume}{39}},
  \bibinfo{pages}{1917} (\bibinfo{year}{2014}{\natexlab{b}}).

\bibitem[{\citenamefont{Ou et~al.}(1991)\citenamefont{Ou, Wang, Zou, and
  Mandel}}]{Wang90}
\bibinfo{author}{\bibfnamefont{Z.}~\bibnamefont{Ou}},
  \bibinfo{author}{\bibfnamefont{L.}~\bibnamefont{Wang}},
  \bibinfo{author}{\bibfnamefont{X.}~\bibnamefont{Zou}}, \bibnamefont{and}
  \bibinfo{author}{\bibfnamefont{L.}~\bibnamefont{Mandel}},
  \bibinfo{journal}{J. Opt. Soc. Am. B} \textbf{\bibinfo{volume}{8}},
  \bibinfo{pages}{978} (\bibinfo{year}{1991}).

\bibitem[{\citenamefont{Wang et~al.}(1991)\citenamefont{Wang, Zou, and
  Mandel}}]{Wang91}
\bibinfo{author}{\bibfnamefont{L.}~\bibnamefont{Wang}},
  \bibinfo{author}{\bibfnamefont{X.}~\bibnamefont{Zou}}, \bibnamefont{and}
  \bibinfo{author}{\bibfnamefont{L.}~\bibnamefont{Mandel}},
  \bibinfo{journal}{J. Opt. Soc. Am. B} \textbf{\bibinfo{volume}{8}},
  \bibinfo{pages}{978} (\bibinfo{year}{1991}).

\bibitem[{\citenamefont{Souto~Ribeiro et~al.}(2001)\citenamefont{Souto~Ribeiro,
  Caetano, Almeida, Huguenin, Coutinho~dos Santos, and Khoury}}]{PHSR2001}
\bibinfo{author}{\bibfnamefont{P.~H.} \bibnamefont{Souto~Ribeiro}},
  \bibinfo{author}{\bibfnamefont{D.~P.} \bibnamefont{Caetano}},
  \bibinfo{author}{\bibfnamefont{M.~P.} \bibnamefont{Almeida}},
  \bibinfo{author}{\bibfnamefont{J.~A.} \bibnamefont{Huguenin}},
  \bibinfo{author}{\bibfnamefont{B.}~\bibnamefont{Coutinho~dos Santos}},
  \bibnamefont{and} \bibinfo{author}{\bibfnamefont{A.~Z.}
  \bibnamefont{Khoury}}, \bibinfo{journal}{Phys. Rev. Lett.}
  \textbf{\bibinfo{volume}{87}}, \bibinfo{pages}{133602}
  (\bibinfo{year}{2001}).

\bibitem[{\citenamefont{Arruda et~al.}(2018)\citenamefont{Arruda, Soares,
  Walborn, Tasca, Kanaan, de~Ara{\'u}jo, and Ribeiro}}]{Arruda2018}
\bibinfo{author}{\bibfnamefont{M.}~\bibnamefont{Arruda}},
  \bibinfo{author}{\bibfnamefont{W.}~\bibnamefont{Soares}},
  \bibinfo{author}{\bibfnamefont{S.}~\bibnamefont{Walborn}},
  \bibinfo{author}{\bibfnamefont{D.}~\bibnamefont{Tasca}},
  \bibinfo{author}{\bibfnamefont{A.}~\bibnamefont{Kanaan}},
  \bibinfo{author}{\bibfnamefont{R.~M.} \bibnamefont{de~Ara{\'u}jo}},
  \bibnamefont{and} \bibinfo{author}{\bibfnamefont{P.~S.}
  \bibnamefont{Ribeiro}}, \bibinfo{journal}{Physical Review A}
  \textbf{\bibinfo{volume}{98}}, \bibinfo{pages}{023850}
  (\bibinfo{year}{2018}).

\bibitem[{\citenamefont{Caetano et~al.}(2002)\citenamefont{Caetano, Almeida,
  Souto~Ribeiro, Huguenin, Coutinho~dos Santos, and Khoury}}]{Caetano2002}
\bibinfo{author}{\bibfnamefont{D.}~\bibnamefont{Caetano}},
  \bibinfo{author}{\bibfnamefont{M.}~\bibnamefont{Almeida}},
  \bibinfo{author}{\bibfnamefont{P.}~\bibnamefont{Souto~Ribeiro}},
  \bibinfo{author}{\bibfnamefont{J.}~\bibnamefont{Huguenin}},
  \bibinfo{author}{\bibfnamefont{B.}~\bibnamefont{Coutinho~dos Santos}},
  \bibnamefont{and} \bibinfo{author}{\bibfnamefont{A.}~\bibnamefont{Khoury}},
  \bibinfo{journal}{Physical Review A} \textbf{\bibinfo{volume}{66}},
  \bibinfo{pages}{041801} (\bibinfo{year}{2002}).

\bibitem[{\citenamefont{de~Oliveira et~al.}(2019)\citenamefont{de~Oliveira,
  Arruda, Soares, Walborn, Khoury, Kanaan, Ribeiro, and
  de~Ara{\'u}jo}}]{Oliveira2019}
\bibinfo{author}{\bibfnamefont{A.~G.} \bibnamefont{de~Oliveira}},
  \bibinfo{author}{\bibfnamefont{M.~F.} \bibnamefont{Arruda}},
  \bibinfo{author}{\bibfnamefont{W.~C.} \bibnamefont{Soares}},
  \bibinfo{author}{\bibfnamefont{S.~P.} \bibnamefont{Walborn}},
  \bibinfo{author}{\bibfnamefont{A.~Z.} \bibnamefont{Khoury}},
  \bibinfo{author}{\bibfnamefont{A.}~\bibnamefont{Kanaan}},
  \bibinfo{author}{\bibfnamefont{P.~S.} \bibnamefont{Ribeiro}},
  \bibnamefont{and} \bibinfo{author}{\bibfnamefont{R.~M.}
  \bibnamefont{de~Ara{\'u}jo}}, \bibinfo{journal}{Brazilian Journal of Physics}
  \textbf{\bibinfo{volume}{49}}, \bibinfo{pages}{10} (\bibinfo{year}{2019}).

\bibitem[{\citenamefont{White et~al.}(1999)\citenamefont{White, James,
  Eberhard, and Kwiat}}]{Kwiat1999}
\bibinfo{author}{\bibfnamefont{A.~G.} \bibnamefont{White}},
  \bibinfo{author}{\bibfnamefont{D.~F.~V.} \bibnamefont{James}},
  \bibinfo{author}{\bibfnamefont{P.~H.} \bibnamefont{Eberhard}},
  \bibnamefont{and} \bibinfo{author}{\bibfnamefont{P.~G.} \bibnamefont{Kwiat}},
  \bibinfo{journal}{Phys. Rev. Lett.} \textbf{\bibinfo{volume}{83}},
  \bibinfo{pages}{3103} (\bibinfo{year}{1999}).

\bibitem[{\citenamefont{Jozsa}(1994)}]{jozsa1994fidelity}
\bibinfo{author}{\bibfnamefont{R.}~\bibnamefont{Jozsa}},
  \bibinfo{journal}{Journal of modern optics} \textbf{\bibinfo{volume}{41}},
  \bibinfo{pages}{2315} (\bibinfo{year}{1994}).

\bibitem[{\citenamefont{Jerrard}(1982)}]{jerrard1982modern}
\bibinfo{author}{\bibfnamefont{H.}~\bibnamefont{Jerrard}},
  \bibinfo{journal}{Optics \& Laser Technology} \textbf{\bibinfo{volume}{14}},
  \bibinfo{pages}{309} (\bibinfo{year}{1982}).

\end{thebibliography}

\end{document}